\documentclass[aps,prd,onecolumn,eqsecnum,amsmath,nofootinbib,preprintnumbers]{revtex4}%

\usepackage[dvips]{color,graphicx}
\usepackage{amsfonts,amssymb,theorem,mathrsfs,times}
\usepackage{bm}
\usepackage[T1]{fontenc} 
\usepackage{comment}
\usepackage{soul}
\usepackage[textsize=footnotesize,textwidth=2.cm]{todonotes}
\usepackage{float}

\newcommand{\ba}{\begin{array}}
\newcommand{\ea}{\end{array}}
\newcommand{\bi}{\begin{itemize}}
\newcommand{\ei}{\end{itemize}}
\newcommand{\bea}{\begin{eqnarray}}
\newcommand{\eea}{\end{eqnarray}}
\newcommand{\be}{\begin{equation}}
\newcommand{\ee}{\end{equation}}

{\theorembodyfont{\upshape}
}
{\theorembodyfont{\upshape}
}
{\theorembodyfont{\upshape}
}
{\theorembodyfont{\upshape}
}
{\theorembodyfont{\upshape}
}
{\theorembodyfont{\upshape}
}

\newcommand{\dalm}{\kern1pt\vbox{\hrule height 0.9pt\hbox{\vrule width
0.9pt\hskip 2.5pt\vbox{\vskip 5.5pt}\hskip 3pt\vrule width 0.3pt}\hrule height
0.3pt}\kern1pt}

\begin{document}
\title{Holographic entanglement entropy in AdS$_3$/WCFT
}

%

\author{Baizhi Gao$^{a,b}$\footnote[2]{e-mail address:
baizhigao@hotmail.com}, Jianfei Xu$^{a,c}$\footnote[1]{e-mail address:
jfxu@seu.edu.cn}}


\affiliation{$^a$ Shing-Tung Yau Center, Southeast University, Nanjing, 210000, China}

\affiliation{$^b$ School of Physics, Southeast University, Nanjing, 211189, China}

\affiliation{$^c$ School of Mathematics, Southeast University, Nanjing, 211189, China}

\date{\today}

\begin{abstract}
In this paper, we consider AdS$_3$ with Comp$\grave{\mathrm{e}}$re-Song-Strominger (CSS) boundary conditions, under which the dual field theory is warped conformal field theory (WCFT), featuring a Virasoro-Kac-Moody algebra. We provide a holographic dual picture of the single interval entanglement entropy in WCFT in the bulk AdS$_3$. We explicitly show that the bulk object captures the entanglement entropy in WCFT is a massive spinning particle's worldline that anchored on the boundary of the interval. Given the mass and spin of the particle as functions of quantum numbers of the twist operator in WCFT, the on-shell worldline action will match the entanglement entropy in WCFT.
\end{abstract}


\maketitle


\section{Introduction}
Holographic duality plays an important role in the recent developments of theoretical physics. The holographic principle is believed to be a fundamental nature lies between gravity theories and quantum field theories. Under the holographic duality, a bulk gravitational physics at weak coupling, which are usually described by geometrical quantities, can be understood from the quantum degrees of freedom in a boundary field theory at strong coupling. This leads to a thought that the spacetime of gravity is an emergent low-energy phenomenon. A recent key progress in this study is the holographic entanglement entropy proposal due to Ryu and Takayanagi~\cite{Ryu:2006bv, Ryu:2006ef}. In the context of AdS$/$CFT, they proposed that the entanglement entropy of a spatial region in the conformal field theory (CFT), which is defined as the von Neumann entropy of the spatial region reduced density matrix, has an elegant bulk formula in Anti de-Sitter space (AdS):
\begin{equation}
S_{\mathrm{EE}}=\frac{A(\gamma)}{4G}\,,
\end{equation}
where $A(\gamma)$ is the area of a bulk minimal surface $\gamma$ anchored on the boundary of the spatial region. This formula provide us a powerful tool to understand the relationship between spacetime geometry and quantum entanglement.

In this paper, we are going to generalize the holographic entanglement entropy proposal to a different class of field theories, namely, warped conformal field theories (WCFTs), which also allow holographic interpretations. Conformal field theories in two spacetime dimensions is a kind of most well-studied quantum field theories. They have infinitely many local symmetries, which can help fix many structures of the underling theories. However, in two spacetime dimensions, this property is not contingent to CFT. It is shown in~\cite{Hofman:2011zj} that a two dimensional field theory with two translational invariance and a chiral scaling symmetry, which indicate a $SL(2, R)\times U(1)$ global symmetry, can have enhanced local symmetry algebra. There are two minimal options for this algebra. One is two copies of the Virasoro algebra which will leads to a CFT, and the other is one copy of Virasoro algebra plus a $U(1)$ Kac-Moody algebra. The later choice will leads to a WCFT. Specific models of WCFT include chiral Liouville gravity~\cite{Compere:2013aya}, free Weyl fermions~\cite{Hofman:2014loa,Castro:2015uaa}, free scalars~\cite{Jensen:2017tnb}, and also the Sachdev-Ye-Kitaev models with complex fermions~\cite{Davison:2016ngz} as a symmetry-broken phase~\cite{Chaturvedi:2018uov}. Due to the infinite symmetries, WCFT is also very constraining. The form of the correlation functions can be fixed without involving a specific model~\cite{Song:2017czq}. Another important motivation for studying such field theories comes from holography for a large class of geometries with $SL(2, R)\times U(1)$ isometry, including the near horizon of extremal Kerr (NHEK)~\cite{Bardeen:1999px, Guica:2010ej} and the warped AdS$_3$ spacetime (WAdS)~\cite{Anninos:2008fx}, and the field theories with $SL(2, R)\times U(1)$ global symmetry having infinite dimensional local symmetries can also be seen from the bulk asymptotic analysis~\cite{Compere:2008cv,Compere:2009zj,Compere:2013bya,Compere:2014bia}. The holographic dualities like Kerr/CFT~\cite{Guica:2008mu}, WAdS/CFT~\cite{Anninos:2008fx}, WAdS/WCFT~\cite{Detournay:2012pc}, and AdS$_3$/WCFT~\cite{Compere:2013bya} are those beyond the standard AdS/CFT correspondence.

We will work in the context of AdS$_3$/WCFT. A Dirichlet-Neumann type boundary conditions, later called the Comp\`{e}re-Song-Strominger (CSS) boundary conditions~\cite{Compere:2013bya} can be imposed to AdS$_3$. Under these conditions, the asymptotic symmetry algebra of AdS$_3$ take a similar form to those in WCFT. This provides an alternative choice of dual field theory for AdS$_3$. WCFTs are not Lorentzian invariant, but share the modular covariance like CFTs. For finite temperature WCFT defined on a torus, it can be shown that the partition functions transform covariantly under the action of modular transformation which exchange the two circles of the torus. This modular property of WCFT was first obtained in~\cite{Detournay:2012pc}, and further explained in~\cite{Castro:2015csg, Song:2019txa}. The density of states at high temperature can be evaluated by using the modular property, which gives us a Cardy like formula for the thermal entropy of WCFT~\cite{Detournay:2012pc}. Holographically, the thermal entropy of WCFT, later called the DHH entropy formula is found to match the black hole entropy of WAdS~\cite{Detournay:2012pc} and of AdS$_3$ with CSS boundary conditions~\cite{Song:2016gtd}. The entanglement entropy of a single interval in WCFT can be calculated by the Rindler method~\cite{Castro:2015csg}. After the Rindler transformation, the causal domain of the interval is mapped to a Rindler space. Then the thermal entropy of the Rindler space can be calculated by a further modular transformation which can reflect the entanglement entropy of that interval. Holographically, a bulk AdS and WAdS picture of the entanglement entropy has been found in~\cite{Song:2016gtd} by translating the Rindler transformation to a quotient in the bulk. Compare to the the Ryu and Takayanagi formula, this bulk picture of the entanglement entropy is still a geodesic as minimal co-dimension two surface in three dimensions, but is not anchored on the end points of the interval. In this work, we are trying to find an alternative bulk picture of the single interval entanglement entropy in WCFT by replacing the geodesic length with a massive spinning particle's worldline action in AdS$_3$. As we will see, our trajectory of the particle can still be chosen as geodesic. In addition to the length part, we add an extrinsic torsion term which reveals the change of the orientation of the spin direction alone the geodesic. With this additional term, the geodesic line is connected to the end points of the interval on the boundary, and the value of the on-shell worldline action of the spinning particle can be shown to match the entanglement entropy in WCFT.

This paper is organized as follows. In section \ref{sec2}, we will give the explanation of the single interval entanglement entropy in WCFT by using the twist operator two point function. In section \ref{sec3}, we study the kinematics of a massive spinning particle on a BTZ black hole background, and we calculated the on-shell worldline action by choosing a specific set of normal vectors. In section \ref{sec4}, we propose that in classical limit, the two point function of spinning operators in field theory can be evaluated by a bulk dual spinning particle's on-shell worldline action, and we match the entanglement entropy of WCFT to a spinning probe's on-shell action with mass and spin all equal to $1/(8G)$.

\section{Entanglement entropy in warped conformal fields theories}\label{sec2}
In a pure field theoretic setup, a two dimensional field theory with translational invariance $x'=x+x_0,~y'=y+y_0$ and a scaling symmetry $x'=\lambda x$ which implies that the global symmetry is $SL(2, R)\times U(1)$ can be shown to have enhanced local symmetry. One minimal choice is the so-called warped conformal symmetry with Virasoro algebra plus a $U(1)$ Kac-Moody algebra as its local algebra~\cite{Hofman:2011zj, Detournay:2012pc}. In position space, a general warped conformal symmetry transformation can be written as
\begin{equation}
x'=f(x),~~~~y'=y+g(x)\,,
\end{equation}
where $x$ and $y$ are $SL(2, R)$ and $U(1)$ local coordinates, and $f(x)$ and $g(x)$ are two arbitrary functions. The above property was later used as a definition of the WCFT~\cite{Detournay:2012pc}. On a cylinder, the conserved charges can be written in terms of Fourier modes, and the later named canonical WCFT algebra takes the following form~\cite{Detournay:2012pc},
\begin{align}\label{wcftalg}
[L_n,L_m]=&(n-m)L_{n+m}+\frac{c}{12}(n^3-n)\delta_{n+m}\,,
\cr
[L_n,P_m]=&-m P_{n+m}\,,
\cr
[P_n,P_m]=&\frac{k}{2}n\delta_{n+m}\,,
\end{align}
Describing a Virasoro algebra with central charge $c$ and a Kac-Moody algebra with level $k$.

Much like its CFT cousin, WCFT data are those spectrum of operators and three point coefficients. The global symmetry of WCFT which is generated by $\{L_{0,\pm1}, P_0\}$ can be used to fix the form of the correlation functions~\cite{Song:2017czq}. The two point function of primary operators is determined by the scaling dimension $\Delta$ which is the eigenvalue of $L_0$ and the $U(1)$ charge $Q$ under the action of $P_0$. The three point function of primary operators further depends on the three point coefficient in addition to the scaling dimension and $U(1)$ charge. For the WCFT entanglement entropy, a Rindler method can be applied which effectively map the causal domain of a subsystem to the Rindler space, and the thermal entropy in Rindler space reflects the entanglement entropy~\cite{Castro:2015csg, Song:2016gtd}.

Due to the constraining of the symmetry group, a replica trick can be applied to the WCFT. The entanglement entropy and R$\acute{\mathrm{e}}$nyi entropy can be calculated by using the two point function of twist operators inserted at the endpoints of the subsystem. On $n$ copy of the original manifold, the trace of the $n-$th power of the reduced density matrix can be evaluated by the two point function of twist operators on a complex plane, so the level $n$ R$\acute{\mathrm{e}}$nyi entropy can be expressed as
\bea\label{renyi}
S_n&=&\frac{1}{1-n}\log\frac{\mathrm{tr}(\rho_{\mathcal{D}}^n)}{(\mathrm{tr}\rho_{\mathcal{D}})^n}\,=\frac{1}{1-n}\log\frac{\langle\Phi_n(x_1, y_1)\Phi_n^{\dagger}(x_2, x_2)\rangle_{\mathcal{C}}}{\langle\Phi_1(x_1, y_1)\Phi_1^{\dagger}(x_2, y_2)\rangle_{\mathcal{C}}^n}\,.
\eea
Here in the first equality, the R$\acute{\mathrm{e}}$nyi entropy is related to the $n$th power of the reduced density matrix $\rho_{\mathcal{D}}$ for a interval $\mathcal{D}$. This can be realized as a path integral on a manifold $\mathcal{R}_n$ which is made up of $n$ decoupled copies of the original space $\mathcal{R}_1$. In the second equality, $\Phi_n$ is the twist operator inserted at the endpoints of the interval that enforce the replica boundary conditions on a plane $\mathcal{C}$. $(x_{1, 2}, y_{1, 2})$ are the endpoint coordinates of the interval $\mathcal{D}$. Given the two point function of the twist operator, one can evaluate the R$\acute{\mathrm{e}}$nyi entropy of the subsystem. The twist operator two point function depends on the conserved charges of $\Phi_n$, and the charges can be further determined by noting that there are two different approaches for evaluating the expectation values of the energy momentum tensor and the $U(1)$ current~\cite{Castro:2015csg, Song:2017czq}. For the twist operators in WCFT, the scaling dimension and the $U(1)$ charge take the following form~\cite{Song:2017czq},
\begin{equation}\label{cdc}
\Delta_n=n\left(\frac{c}{24}+\frac{L_0^{vac}}{n^2}+\frac{iP_0^{vac}\alpha}{2n\pi}-\frac{\alpha^2k}{16\pi^2}\right),~~~~Q_n=n\left(-\frac{P_0^{vac}}{n}-i\frac{k\alpha}{4\pi}\right)\,,
\end{equation}
where $L_0^{vac}$ and $P_0^{vac}$ are the vacuum expectation values of $L_0$ and $P_0$ respectively, and $\alpha$ is a constant which is related to the spectral flow parameter of the WCFT algebra~\cite{Apolo:2018oqv}. Given the charges above, and the form of the two point function of the twist operators in WCFT~\cite{Song:2017czq},
\begin{equation}\label{2pf}
\langle\Phi_n(x_1, y_1)\Phi_n^{\dagger}(x_2, y_2)\rangle_{\mathcal{C}}\sim e^{iQ_n\left(y_1-y_2+\frac{\bar{\beta}-\alpha}{\beta}(x_1-x_2)\right)}\left(\frac{\beta}{\pi}\sinh\frac{\pi(y_1-y_2)}{\beta}\right)^{-2\Delta_n}\,.
\end{equation}
where $\beta$ and $\bar{\beta}$ are the inverse temperature along $x$ and $y$, the R$\acute{\mathrm{e}}$nyi entropy as a function of such constants and the interval can be evaluated,
\begin{eqnarray}
S_n&=&\frac{1}{1-n}\log\frac{\mathrm{tr}(\rho_{\mathcal{D}}^n)}{(\mathrm{tr}\rho_{\mathcal{D}})^n}=\frac{1}{1-n}\log\frac{\langle\Phi_n(x_1, y_1)\Phi_n^{\dagger}(x_2, y_2)\rangle_{\mathcal{C}}}{\langle\Phi_1(x_1, y_1)\Phi_1^{\dagger}(x_2, y_2)\rangle_{\mathcal{C}}^n}\nonumber\\
&=&-iP_0^{vac}\left(\Delta y+\frac{\bar{\beta}-\alpha}{\beta}\Delta x\right)+\left(-\frac{\alpha}{\pi}iP_0^{vac}-\frac{2(n+1)}{n}L_0^{vac}\right)\log\left(\frac{\beta}{\pi}\sinh\frac{\pi\Delta x}{\beta}\right)\,,\nonumber\\\label{eefield}
\end{eqnarray}
where $\Delta x=x_2-x_1$, $\Delta y=y_2-y_1$. The entanglement entropy is a special case when $n=1$ in the R$\acute{\mathrm{e}}$nyi entropy. In fact, the dependence of the subsystem interval in the expression of the entanglement entropy $S_{EE}=S_1$ is understandable. For the WCFT, the $SL(2, R)$ part shares the scaling invariance which leads to a logarithmic dependence, while the $U(1)$ part with a spectral flow is a nonlocal direction which leads to a linear dependence.

In the limit $n\to1$, and using equation \eqref{renyi}, the entanglement entropy can be written as
\begin{equation}\label{ee}
S_{\mathrm{EE}}=\lim_{n\to1}S_n=-\left(\partial_n\log\langle\Phi_n(x_1, y_1)\Phi_n^{\dagger}(x_2, y_2)\rangle_{\mathcal{C}}-\log\langle\Phi_1(x_1, y_1)\Phi_1^{\dagger}(x_2, y_2)\rangle_{\mathcal{C}}\right)|_{n=1}\,.
\end{equation}
In the following sections, we will find out a holographic dual of the twist operator two point function in the bulk AdS$_3$ as a spinning particle's on-shell action, so the holographic picture of the entanglement entropy in WCFT can be specified by a worldline of a spinning particle with end points anchored on the boundary. The reason why we consider spinning particle is that under the action of the warped conformal group, the twist operator's spin angular momentum is nonzero, so it's natural to consider a massive spinning particle as its dual object. This will be clear in the following sections.

\section{Massive spinning particles in BTZ black hole spacetime}\label{sec3}
The BTZ black hole spacetime metric can be written in a form with light-like coordinates,
\begin{equation}\label{BTZ}
ds^2=T_u^2du^2+2\rho dudv+T_v^2dv^2+\frac{d\rho^2}{4(\rho^2-T_uT_v)}\,,
\end{equation}
along with the identifications
\begin{equation}\label{id}
u\sim u+2\pi,~~~~v\sim v+2\pi\,.
\end{equation}
$T_u$ and $T_v$ are variable constants. The local isometry for BTZ black hole is $SL(2, R)\times SL(2, R)$, while only the $U(1)\times U(1)$ part is globally defined due to the spatial circle.

For a particle with mass $m$ and spin $s$, the worldline action can be written as~\cite{Castro:2014tta}
\begin{equation}\label{wac}
S=\int d\tau\left(m\sqrt{g_{\mu\nu}\dot{X}^{\mu}\dot{X}^{\nu}}+s\tilde{n}\cdot\nabla n\right)+S_{constraints}\,.
\end{equation}
Here $\tau$ is the length parameter, $\dot{X}^{\mu}$ is the tangent vector of the worldline, and $S_{constraints}$ contains Lagrange multipliers which require that the two normalized vectors $n$ and $\tilde{n}$ should be mutually orthogonal and perpendicular to the worldline, i.e.,
\begin{equation}\label{constraint}
n^2=-1,~~~~\tilde{n}^2=1,~~~~n\cdot\tilde{n}=0,~~~~n\cdot\dot{X}=\tilde{n}\cdot\dot{X}=0\,.
\end{equation}
The symbol $\nabla$ with no subscript indicates a covariant derivative alone the worldline:
\begin{equation}
\nabla V^{\mu}:=\dot{X}^{\nu}\nabla_{\nu}V^{\mu}\,.
\end{equation}
The equations of motion with respect to $X^{\mu}(\tau)$ are known as the Mathisson-Papapetrou-Dixon (MPD) equations,
\begin{equation}\label{MPD}
\nabla[m\dot{X}^{\mu}+\dot{X}^{\nu}\nabla s^{\mu}{}_{\nu}]=-\frac{1}{2}\dot{X}^{\nu}s^{\rho\sigma}R^{\mu}{}_{\nu\rho\sigma}\,,
\end{equation}
where $s^{\mu\nu}$ is the spin tensor,
\begin{equation}
s^{\mu\nu}=s(n^{\mu}\tilde{n}^{\nu}-\tilde{n}^{\mu}n^{\nu})\,.
\end{equation}
As was noticed in~\cite{Castro:2014tta}, in locally AdS spacetimes, the contraction of the Riemann tensor with $s^{\mu\nu}\dot{X}^{\rho}$ vanishes. The MPD equations reduce to
\begin{equation}\label{MPD0}
\nabla[m\dot{X}^{\mu}-s^{\mu}{}_{\nu}\nabla\dot{X}^{\nu}]=0\,.
\end{equation}
Although the worldline action \eqref{wac} is obtained in~\cite{Castro:2014tta} from the topological massive gravity by using the conical method developed in~\cite{Lewkowycz:2013nqa}, here we use it as a known result for spinning particle in three dimensional Einstein gravity, since the MPD equation \eqref{MPD} works geometrically in any background. One obvious solution to the MPD equation above is a geodesic,
\begin{equation}\label{geod}
\nabla\dot{X}^{\mu}=0\,.
\end{equation}
We are trying to find the path that minimise the worldline action \eqref{wac}. However, the local minimum or the saddle point of the worldline action is not necessary a geodesic. Here we will take the solution to MPD equation in AdS$_3$ as a geodesic. The first reason is that it is the simplest case for discussing geometric quantities in the bulk. The second reason comes from the fact that in the Frenet-Serret gauge with one of the extrinsic curvatures of a path vanishes identically~\cite{Fonda:2018eqg}, the geodesic one takes a smaller value of the on-shell worldline action. In~\cite{Fonda:2018ctf}, the relationship between the entanglement entropy of a CFT with gravitational anomaly and the non-geodesic Mathisson's helices has been studied. We will left the 
discussions of the non-geodesic case when matching the WCFT entanglement entroy in our future works.

We first consider a massive spinning particle moving in the BTZ black hole with zero temperature, i.e., $T_{u, v}=0$. We choose the solution to the MPD equation as geodesic, and we set the starting and ending point of the particle all fixed on the AdS$_3$ boundary. The geodesic line that connect $\{-\frac{\Delta u}{2}, -\frac{\Delta v}{2}, \infty\}$ and $\{\frac{\Delta u}{2}, \frac{\Delta v}{2}, \infty\}$ can be parameterized as
\begin{eqnarray}
u(\tau)&=&\frac{\Delta u}{2}\tanh\left(\tau+\frac{1}{2}\log(\Delta u\Delta v)\right)\,,\\
v(\tau)&=&\frac{\Delta v}{2}\tanh\left(\tau+\frac{1}{2}\log(\Delta u\Delta v)\right)\,,\\
\rho(\tau)&=&\frac{1}{2}\left(e^\tau+\frac{1}{\Delta u\Delta v}e^{-\tau}\right)^2\,.\label{radialcoord}
\end{eqnarray}
where $\tau$ parameterise the length of the geodesic varying from $-\infty$ to $+\infty$. The tangent vector of the geodesic can be written down as derivative with respect to $\tau$,
\begin{equation}\label{tv}
\dot{X}^{\mu}\partial_{\mu}=\frac{1/\Delta v}{\rho}\partial_{u}+\frac{1/\Delta u}{\rho}\partial_{v}+\frac{4u\rho}{\Delta u}\partial_{\rho}.
\end{equation}

On-shelly, the worldline action consist of two parts. The first part is the length of the geodesic times its mass, and the second part is the extrinsic torsion term due to the spin.
\begin{equation}\label{onshellaction}
S_{\mathrm{on-shell}}=mL_{\mathrm{geo}}+S_{\mathrm{torsion}}=mL_{\mathrm{geo}}+s\int d\tau\tilde{n}\cdot\nabla n\,.
\end{equation}
If we label the radial coordinate of the geodesic close to the boundary as $\rho_{\infty}$, the length parameter of the two end points can be read off from \eqref{radialcoord},
\begin{equation}
\tau_f=\log\sqrt{2\rho_{\infty}},~~~~\tau_i=\log\frac{1}{\sqrt{2\rho_{\infty}}\Delta u\Delta v}\,,
\end{equation}
then the length of the geodesic is the difference of $\tau_f$ and $\tau_i$, and this is actually divergent as expected according to the infinity of $\rho_{\infty}$,
\begin{equation}\label{length}
L_{\mathrm{geo}}=\tau_f-\tau_i=\log\frac{\Delta u\Delta v}{1/(2\rho_{\infty})}\,.
\end{equation}
The inverse of $2\rho_{\infty}$ can be viewed as a UV cutoff in the u-v plane.

The worldline of the particle as a one dimensional object in the bulk has intrinsic and extrinsic properties, which can help to fix the geometrical nature of the worldline. The intrinsic properties can be obtained from the tangent vector $\dot{X}^{\mu}$, and its normalization $\dot{X}^{\mu}\dot{X}_{\mu}=1$ indicates that the worldline has trivial induced metric which only characterize the reparameterization of that line. The extrinsic quantities can be figured out by studying how the tangent and normal vectors changes as we move alone the worldline. One can always decompose the tangent directional derivative $\nabla$ of the normal vectors into their tangent and normal contributions,
\begin{eqnarray}
\nabla n^{\mu}&=&K^n\dot{X}^{\mu}-T^{n\tilde{n}}\tilde{n}^{\mu}\,,\\
\nabla\tilde{n}^{\mu}&=&K^{\tilde{n}}\dot{X}^{\mu}-T^{\tilde{n}n}n^{\mu}\,,
\end{eqnarray}
where $K$ and $T$ tensors are the extrinsic curvature and extrinsic torsion respectively. The geodesic line has vanishing extrinsic curvature, i.e., $K^{n}=-n\cdot\nabla\dot{X}=0$, $K^{\tilde{n}}=-\tilde{n}\cdot\nabla\dot{X}=0$ due to the geodesic equation \eqref{geod}.

The extrinsic torsion tensor can be expressed as $T^{n\tilde{n}}=-T^{\tilde{n}n}=-\tilde{n}\cdot\nabla n$, which is exactly the integrand of the second term of the on-shell action \eqref{onshellaction}, characterizing the twist of the worldline. When a particle has spin, the spin direction can be used to determine a local frame $(q, \tilde{q})$, with $q^2=-1, \tilde{q}^2=1$, which is parallelly transported along its trajectory, i.e. $\nabla q(\tilde{q})=0$. If the worldline is spacelike, the integral of the extrinsic torsion along that line is determined by the change of the rapidity of the Lorentz boost from parallelly transported frame to the normal frame between the two end points~\cite{Castro:2014tta},
\begin{equation}\label{anglechange}
S_{\mathrm{torsion}}=s(\eta_f-\eta_i)\,,
\end{equation}
where $\eta$ is the rapidity which is the Lorentz angle between $(n, \tilde{n})$ and $(q, \tilde{q})$. This change of the rapidity or the extrinsic torsion term can always be chosen as a positive value, since it reveals the rotation of the spin direction from initial to final. In fact, given any two normal frames that belong to the same worldline, there exist a local Lorentz transformation that relates these two normal frames on each point on that line. The difference of the two torsions defined by these two normal frames is just a line derivative of the local angle parametrizing the local rotation between them~\cite{Fonda:2018eqg}. So the integral of the extrinsic torsion along the worldline is invariant under local rotation of the normal frame up to boundary contribution from the end points, and its value depends on the specific choice of the local frame.

Here we give our recipe for choosing the normal vectors. Given a parallelly transported frame $(q, \tilde{q})$ along the trajectory, there exists two sets of normal vectors $(n_1, \tilde{n}_1)$ and $(n_2, \tilde{n}_2)$, which have opposite Lorentz angle to $(q, \tilde{q})$ at each point. Set a time direction on the boundary, and take one of the time-like normal vectors alone that time on the boundary, one of the two sets of normal vectors will give out a extrinsic torsion term that is responsible for the entanglement entropy in CFT with gravitational anomaly~\cite{Castro:2014tta} and the other corresponds to WCFT. Geometrically, the normal vectors $n$ and $\tilde{n}$ are perpendicular to the geodesic and so to $\dot{X}$. We choose our normal frame with the normal vectors satisfying \eqref{constraint} as the following,
\begin{eqnarray}
n^{\mu}\partial_{\mu}&=&\pm\frac{\Delta u^2\sqrt{\rho^2-\frac{2\rho}{\Delta u\Delta v}}}{\rho\sqrt{2\Delta u^2\Delta v^2\rho-(\Delta u+\Delta v)^2}}\partial_u\mp\frac{\Delta v^2\sqrt{\rho^2-\frac{2\rho}{\Delta u\Delta v}}}{\rho\sqrt{2\Delta u^2\Delta v^2\rho-(\Delta u+\Delta v)^2}}\partial_v\nonumber\\
&&+\frac{2\rho(\Delta u-\Delta v)}{\sqrt{2\Delta u^2\Delta v^2\rho-(\Delta u+\Delta v)^2}}\partial_{\rho}\,,\label{nvn}\\
\tilde{n}^{\mu}\partial_{\mu}&=&\frac{\Delta u+\Delta v-\Delta u^2\Delta v\rho}{\Delta v\rho\sqrt{2\Delta u^2\Delta v^2\rho-(\Delta u+\Delta v)^2}}\partial_u+\frac{\Delta u+\Delta v-\Delta u\Delta v^2\rho}{\Delta u\rho\sqrt{2\Delta u^2\Delta v^2\rho-(\Delta u+\Delta v)^2}}\partial_v\nonumber\\\
&&\pm\frac{2(\Delta u+\Delta v)\sqrt{\rho^2-\frac{2\rho}{\Delta u\Delta v}}}{\sqrt{2\Delta u^2\Delta v^2\rho-(\Delta u+\Delta v)^2}}\partial_{\rho}\,,\label{nvnt}
\end{eqnarray}
where the up sign corresponds to the left part of the geodesic with $u<0$, and the down sign corresponds to the right part of the geodesic with $u>0$. When $u=0$ or at the turning point of the geodesic with $\tau_m=\frac{1}{2}\log\frac{1}{\Delta u\Delta v}$, we have $\rho=\rho_m=\frac{2}{\Delta u\Delta v}$, and these two sets of normal vectors are smoothly connected. On the boundary, at the end points of the geodesic, the normal vectors take the expressions,
\begin{eqnarray}\label{nb}
n_b^{\mu}\partial_{\mu}&=&\pm\frac{\Delta u}{\Delta v\sqrt{2\rho_{\infty}}}\partial_u\mp\frac{\Delta v}{\Delta u\sqrt{2\rho_{\infty}}}\partial_v\,,\\
\tilde{n}_b^{\mu}\partial_{\mu}&=&-\frac{\Delta u}{\Delta v\sqrt{2\rho_{\infty}}}\partial_u-\frac{\Delta v}{\Delta u\sqrt{2\rho_{\infty}}}\partial_v\,,\label{ntb}
\end{eqnarray}
These boundary values of the normal vectors is a set of our gauge choice of the local Lorentz rotation on the frame. In fact these boundary values together with \eqref{constraint} can uniquely determine \eqref{nvn} and \eqref{nvnt}. The parallelly transported vector $q$ at the end points can also be carried out,
\begin{equation}
q_b^{\mu}\partial_{\mu}=\pm\sqrt{\frac{\Delta u}{2\Delta v\rho_{\infty}}}\partial_u\mp\sqrt{\frac{\Delta v}{2\Delta u\rho_{\infty}}}\partial_v\,,
\end{equation}
where the up sign corresponds to the initial point at $\tau=\tau_i$, and the down sign corresponds to the final point at $\tau=\tau_f$. For the light-cone coordinates $u$ and $v$, the time $t$ and spatial $\sigma$ coordinates can be given by $u=\sigma+t$, $v=\sigma-t$. Here the normal vector, say $n^{\mu}$, we choose is not the simply along $t$. Explicitly, on the boundary, the normalized vector along $t$ takes the form,
\begin{equation}
n_t^{\mu}\partial_{\mu}=\pm\frac{1}{\sqrt{2\rho_{\infty}}}\partial_u\mp\frac{1}{\sqrt{2\rho_{\infty}}}\partial_v\,.
\end{equation}
It can be shown that the Lorentz angle between $n_t$ and $q_b$ is opposite to the Lorentz angle between $n_b$ and $q_b$, which means that $n_t$ and $n_b$ are oppositely boosted. For the choice of $n_t$ as boundary value of the time-like normal vector, the calculations of the extrinsic torsion term in AdS$_3$ have been done in~\cite{Castro:2014tta}, where the dual field theory is taken to be CFT with gravitational anomaly. Here we choose $n_b$ as the time-like vector on the boundary, and try to explain the entanglement entropy in the WCFT. We will see from the following finite temperature case that the normal vectors we choose are consistent with the CSS boundary conditions.

Given the explicit expressions of the normal vectors \eqref{nvn} and \eqref{nvnt}, and the tangent vector \eqref{tv}, we can calculate the extrinsic torsion at each point on the geodesic,
\begin{equation}
\tilde{n}\cdot\nabla n=\frac{\Delta v^2-\Delta u^2}{2\Delta u^2\Delta v^2\rho-(\Delta u+\Delta v)^2}\,.
\end{equation}
Substituting \eqref{radialcoord} into the above expression and do the $\tau$ integration from $\tau_i$ to $\tau_f$, we find,
\begin{equation}\label{torsion}
\int_{\tau_i}^{\tau_f}d\tau\tilde{n}\nabla n=\log\left(\frac{\Delta u}{\Delta v}\right)\,.
\end{equation}

Now we can combine the two parts of the on-shell action together. Given \eqref{length} and \eqref{torsion}, the on-shell worldline action \eqref{onshellaction} can be evaluated,
\begin{equation}\label{oa0}
S_{\mathrm{on-shell}}=m\log\frac{\Delta u\Delta v}{\epsilon}+s\log\frac{\Delta u}{\Delta v}\,,
\end{equation}
where $\epsilon=1/(2\rho_{\infty})$ is the UV cutoff.

For finite temperature, let us consider the case of $T_u=0$ and $T_v$ left as a variable constant. The geodesic line that followed by a massive spinning particle, connecting $\{-\frac{\Delta u}{2}, -\frac{\Delta v}{2}, \infty\}$ and $\{\frac{\Delta u}{2}, \frac{\Delta v}{2}, \infty\}$ can be parameterized as
\begin{eqnarray}
u(\tau)&=&\frac{\Delta u\sinh\left(2\tau+\log\frac{\Delta u\sinh(T_v\Delta v)}{T_v}\right)}{2\left(\cosh(T_v\Delta v)+\cosh\left(2\tau+\log\frac{\Delta u\sinh(T_v\Delta v)}{T_v}\right)\right)}\,,\\
v(\tau)&=&\frac{1}{2T_v}\log\frac{T_v\left(\sinh\left(2\tau+\log\frac{\Delta u\sinh(T_v\Delta v)}{T_v}\right)-\sinh(T_v\Delta v)\right)}{\sinh(T_v\Delta v)\sinh\left(2\tau+\log\frac{\Delta u\sinh(T_v\Delta v)}{T_v}-T_v\Delta v\right)}\,,\\
\rho(\tau)&=&\frac{1}{2}\left(e^{2\tau}+\frac{2T_v\cosh(T_v\Delta v)}{\Delta u\sinh(T_v\Delta v)}+\left(\frac{T_v}{\Delta u\sinh(T_v\Delta v)}\right)^2e^{-2\tau}\right)\,,\label{rhoft}
\end{eqnarray}
where $\tau$ is the length parameter varying from $\tau_i$ to $\tau_f$ with the following expression,
\begin{equation}
\tau_f=\log\sqrt{2\rho_{\infty}},~~~~\tau_i=\log\frac{T_v}{\sqrt{2\rho_{\infty}}\Delta u\sinh(T_v\Delta v)}\,,
\end{equation}
$\rho_{\infty}$ is the radial coordinate near the boundary. The length of the geodesic, which is the difference of the end point parameters, in this case can be written as
\begin{equation}\label{lengthft}
L_{\mathrm{geo}}=\log\frac{\Delta u\sinh(T_v\Delta v)}{(1/(2\rho_{\infty}))T_v}\,.
\end{equation}
The tangent vector of the geodesic takes the following form,
\begin{equation}\label{dX}
\dot{X}^{\mu}\partial_{\mu}=\left(\frac{T_v\coth(T_v\Delta v)}{\rho}-\frac{T_v^2}{\Delta u\rho^2}\right)\partial_u+\frac{1/\Delta u}{\rho}\partial_v+\frac{4u\rho}{\Delta u}\partial_{\rho}\,.
\end{equation}
The normal vectors satisfying \eqref{constraint} in the finite temperature case we choose are the following,
\begin{gather}
n^{\mu}=\begin{pmatrix}\label{nvnft}
\frac{\pm\Delta u \sqrt{\rho ^2+\frac{T_v^2}{\Delta u^2}-\frac{2 \rho  T_v \coth (T_v\Delta v)}{\Delta u}} \left(\frac{T_v^2}{\Delta u^2}+\rho  (T_v \coth (T_v\Delta v)+T_v)^2\right)}{\rho ^2 \sqrt{\frac{T_v^2-T_v^2 \coth ^2(T_v\Delta v)}{\Delta u^2}+(T_v+T_v \coth (T_v\Delta v))^2 \left(2 \rho -\frac{2 T_v \coth (T_v\Delta v)}{\Delta u}-(T_v+T_v \coth (T_v\Delta v))^2\right)}}\,,\\
\frac{\mp\sqrt{\rho ^2+\frac{T_v^2}{\Delta u^2}-\frac{2 \rho  T_v \coth (T_v\Delta v)}{\Delta u}}}{\Delta u\rho\sqrt{\frac{T_v^2-T_v^2 \coth ^2(T_v\Delta v)}{\Delta u^2}+(T_v+T_v \coth (T_v\Delta v))^2 \left(2 \rho -\frac{2 T_v \coth (T_v\Delta v)}{\Delta u}-(T_v+T_v \coth (T_v\Delta v))^2\right)}}\,,\\
\frac{2\left(\frac{T_v^2}{\Delta u^2}-\frac{\rho T_v\coth(T_v\Delta v)}{\Delta u}+\rho  (T_v \coth (T_v\Delta v)+T_v)^2\right)}{\sqrt{\frac{T_v^2-T_v^2 \coth ^2(T_v\Delta v)}{\Delta u^2}+(T_v+T_v \coth (T_v\Delta v))^2 \left(2 \rho -\frac{2 T_v \coth (T_v\Delta v)}{\Delta u}-(T_v+T_v \coth (T_v\Delta v))^2\right)}}
\end{pmatrix}
\end{gather}
\begin{gather}
\tilde{n}^{\mu}=\begin{pmatrix}\label{nvntft}
\frac{-\left(\frac{T_v^2}{\Delta u}+\rho^2\Delta u-\rho T_v\coth(T_v\Delta v)\right)(T_v+T_v\coth(T_v\Delta v))^2+\frac{T_v^3\coth(T_v\Delta v)}{\Delta u^2}-\frac{\rho T_v^2\coth^2(T_v\Delta v)}{\Delta u}}{\rho^2\sqrt{\frac{T_v^2-T_v^2 \coth ^2(T_v\Delta v)}{\Delta u^2}+(T_v+T_v \coth (T_v\Delta v))^2 \left(2 \rho -\frac{2 T_v \coth (T_v\Delta v)}{\Delta u}-(T_v+T_v \coth (T_v\Delta v))^2\right)}}\,,\\
\frac{-\left(\rho-\frac{T_v\coth(T_v\Delta v)}{\Delta u}\right)+(T_v+T_v\coth(T_v\Delta v))^2}{\Delta u\rho\sqrt{\frac{T_v^2-T_v^2 \coth ^2(T_v\Delta v)}{\Delta u^2}+(T_v+T_v \coth (T_v\Delta v))^2 \left(2 \rho -\frac{2 T_v \coth (T_v\Delta v)}{\Delta u}-(T_v+T_v \coth (T_v\Delta v))^2\right)}}\,,\\
\frac{\mp2\sqrt{\rho ^2+\frac{T_v^2}{\Delta u^2}-\frac{2 \rho  T_v \coth (T_v\Delta v)}{\Delta u}}\left(\frac{T_v\coth(T_v\Delta v)}{\Delta u}+(T_v+T_v\coth(T_v\Delta v))^2\right)}{\sqrt{\frac{T_v^2-T_v^2 \coth ^2(T_v\Delta v)}{\Delta u^2}+(T_v+T_v \coth (T_v\Delta v))^2 \left(2 \rho -\frac{2 T_v \coth (T_v\Delta v)}{\Delta u}-(T_v+T_v \coth (T_v\Delta v))^2\right)}}
\end{pmatrix}
\end{gather}
where we use the convention $(u, v, \rho)$ to label the component. The up sign corresponds to the left part of the geodesic with $u<0$, and the down sign corresponds to the right part of the geodesic with $u>0$. These two parts are smoothly connected at the turning point with $\rho_m=\frac{T_v\coth((T_v\Delta)v/2)}{\Delta u}$. The boundary values of these normal vectors are,
\begin{eqnarray}
n_b^{\mu}\partial_{\mu}&=&\pm\frac{\Delta u(T_v+T_v\coth(T_v\Delta v))}{\sqrt{2\rho_{\infty}}}\partial_u\mp\frac{1/(\Delta u(T_v+T_v\coth(T_v\Delta v)))}{\sqrt{2\rho_{\infty}}}\partial_v\,,\\
\tilde{n}_b^{\mu}\partial_{\mu}&=&-\frac{\Delta u(T_v+T_v\coth(T_v\Delta v))}{\sqrt{2\rho_{\infty}}}\partial_u-\frac{1/(\Delta u(T_v+T_v\coth(T_v\Delta v)))}{\sqrt{2\rho_{\infty}}}\partial_v\,,
\end{eqnarray}
where the up sign corresponds to the initial point at $\tau=\tau_i$, and the down sign corresponds to the final point at $\tau=\tau_f$. When $T_v=0$, these vectors recover \eqref{nb} and \eqref{ntb}. 

In the Fefferman-Graham expansion, any three dimensional metric takes the form,
\begin{equation}
ds^2=\frac{d\rho^2}{4\rho^2}+\rho\left(g_{ab}^{(0)}+\frac{1}{\rho}g_{ab}^{(2)}+\mathcal{O}(\rho^{-2})\right)dx^adx^b\,,
\end{equation}
where we use the radial coordinate $\rho=r^2/2$ instead of $r$ in~\cite{Compere:2013bya} and set $\ell=1$. The CSS boundary conditions imposed on the above expansion are,
\begin{eqnarray}\label{CSS}
g_{vv}^{(0)}&=&0,~~~~g_{uu}^{(0)}=\partial_u\bar{P}(u),~~~~g_{uv}^{(0)}=g_{vu}^{(0)}=1\,,\nonumber\\
g_{vv}^{(2)}&=&4G\Delta\,.
\end{eqnarray}
We can use the above normal vectors \eqref{nvnft} and \eqref{nvntft} as well as the tangent vector along the geodesic \eqref{dX} to recover the boundary behavior of the metric components, 
\begin{equation}
g_{\mu\nu}=-n_{\mu}n_{\nu}+\tilde{n}_{\mu}\tilde{n}_{\nu}+\dot{X}_{\mu}\dot{X}_{\nu}\,.
\end{equation}
At large radius $\rho$, the fall of behavior of the metric components can be figures out,
\begin{eqnarray}\label{gex}
g_{\rho\rho}&=&\frac{1}{4\rho^2}+\mathcal{O}(\rho^{-3})\,,\nonumber\\
g_{\rho u}&=&\mathcal{O}(\rho^{-3})\,,~~~g_{\rho v}=\mathcal{O}(\rho^{-3})\,,\nonumber\\
g_{uv}&=&\rho+\frac{\coth^2(T_v\Delta v)}{\Delta u^2(1+\coth(T_v\Delta v))^2}+\mathcal{O}(\rho^{-1})\\
g_{uu}&=&\mathcal{O}(\rho^{-3})\,,\nonumber\\
g_{vv}&=&T_v^2(1+2\coth^2(T_v\Delta v))+\mathcal{O}(\rho^{-1})\,.\nonumber
\end{eqnarray}
We see that the asymptotic expansions \eqref{gex} of the metric components are consistent with the CSS boundary conditions \eqref{CSS}. 

The extrinsic torsion that relates to \eqref{nvnft} and \eqref{nvntft} can be calculated,
\begin{equation}
\tilde{n}\nabla n=\frac{\frac{1}{\Delta u^2}-\frac{T_v^2e^{4T_v\Delta v}}{\sinh^2(T_v\Delta v)}}{2\rho e^{2T_v\Delta v}-\frac{1}{\Delta u^2}-\frac{2T_ve^{2T_v\Delta v}\coth(T_v\Delta v)}{\Delta u}-\frac{T_v^2e^{4T_v\Delta v}}{\sinh^2(T_v\Delta v)}}\,.
\end{equation}
Taking into account the radius function \eqref{rhoft} along the geodesic, the integral of the extrinsic torsion from $\tau_i$ to $\tau_{f}$ is,
\begin{equation}\label{torsionft}
\int_{\tau_i}^{\tau_f}d\tau\tilde{n}\nabla n=\log\frac{T_v\Delta ue^{2T_v\Delta v}}{\sinh(T_v\Delta v)}\,.
\end{equation}
Combing the length \eqref{lengthft} and torsion \eqref{torsionft} parts, the on-shell action of the spinning particle in the finite temperature case can be written as
\begin{equation}\label{oa}
S_{\mathrm{on-shell}}=m\log\frac{\Delta u\sinh(T_v\Delta v)}{\epsilon T_v}+s\log\frac{T_v\Delta ue^{2T_v\Delta v}}{\sinh(T_v\Delta v)}\,,
\end{equation}
where $\epsilon=1/(2\rho_{\infty})$ is the UV cutoff.

\section{Entanglement entropy in AdS$_3$/WCFT}\label{sec4}
Under CSS boundary conditions~\cite{Compere:2013bya}, which is a Dirichlet-Nuemann type boundary condition, asymptotically AdS$_3$ spacetimes are conjectured to be dual to WCFTs in the sense that the symmetry algebras on both sides take a similar form. In the light-like coordinate system, the BTZ black hole metric (\ref{BTZ}) with identification (\ref{id}) satisfies the CSS boundary conditions,
\begin{equation}\label{CSSbc}
g_{uv}^{(0)}=1,~~~g_{vv}^{(0)}=0,~~~\partial_vg_{uu}^{(0)}=0,~~~g_{vv}^{(2)}=T_v^2\,,
\end{equation}
with $T_v$ fixed.
The asymptotic Killing vectors obeying the boundary condition~\eqref{CSSbc} are,
\begin{equation}
\xi_n=e^{inu}\left(\partial_u-\frac{1}{2}in\partial_{\rho}\right),~~~~\eta_n=-e^{inu}\partial_v\,,
\end{equation}
The asymptotic symmetry algebra under the above mentioned boundary conditions is the Virasoro-Kac-Moody algebra, which can be written as
\begin{eqnarray}\label{alg}
{}[\tilde{L}_n, \tilde{L}_m]&=&(n-m)\tilde{L}_{n+m}+\frac{c}{12}(n^3-n)\delta_{n,-m},\nonumber\\
{}[\tilde{L}_n, \tilde{P}_m]&=&-m\tilde{P}_{n+m}+m\tilde{P}_0\delta_{n,-m},\\
{}[\tilde{P}_n, \tilde{P}_m]&=&\frac{\tilde{k}}{2}n\delta_{n,-m}\,,\nonumber
\end{eqnarray}
where
\begin{equation}
c=\frac{3}{2G},~~~~\tilde{k}=-\frac{T_v^2}{G}\,.
\end{equation}
$\tilde{L}_0$ and $\tilde{P_0}$ are the conserved charges associated with left and right moving Killing vectors $\partial_u$ and $\partial_v$ respectively, and are related to the bulk mass and energy momentum by
$\langle\tilde{L}_0\rangle ={1\over2}(E+J),~\langle\tilde{P}_0\rangle=-{1\over2}(E-J)$. The asymptotic charges $\tilde{ L}_n, \tilde{ P}_n$ are both finite and integrable with fixed $T_v$. The above algebra \eqref{alg} is not the canonical WCFT algebra\eqref{wcftalg}. One way to relate them is utilizing the following charge redefinition~\cite{Detournay:2012pc},
\begin{equation}\label{dhhmap}
\tilde{L}_n=L_n-\frac{2P_0P_n}{k}+\frac{P_0^2\delta_{n,0}}{k},~~~~\tilde{P}_n=\frac{2P_0P_n}{k}-\frac{P_0^2\delta_{n,0}}{k}\,.
\end{equation}
In this paper, we are interested in states with $\langle P_n\rangle=0,~\forall n\neq0$, and this amounts to a nonlocal reparameterization of the theory,
\begin{equation}\label{uvxy}
u=x,~~~~v=\frac{ky}{2\langle P_0\rangle}+x\,.
\end{equation}
On such states we further have the expectation values,
\begin{equation}
\langle L_0\rangle=\langle\tilde{L}_0\rangle+\langle\tilde{P}_0\rangle=J,~~~~\frac{\langle P_0\rangle^2}{k}=\langle\tilde{P}_0\rangle=\frac{1}{2}(J-E)\,.
\end{equation}
For the BTZ black holes (\ref{BTZ}), the expectation values of the zero modes of the canonical WCFT algebra (\ref{wcftalg}) can now be read from the background geometry,
\begin{equation}\label{L0P0}
\langle L_0\rangle=\frac{1}{4G}(T_u^2-T_v^2),~~~~\langle P_0\rangle=-\frac{T_v}{2}\sqrt{\frac{-k}{G}}\,.
\end{equation}
The WCFT vacuum corresponds to the global AdS$_3$ with $T_u=T_v=-\frac{i}{2}$, so the vacuum charges can be written down,
\begin{equation}\label{vaccharge}
L_0^{vac}=0,~~~~P_0^{vac}=\frac{i}{4}\sqrt{\frac{-k}{G}}\,.
\end{equation}
The Kac-Moody level $k$ is a free parameter introduced in the charge redefinition \eqref{dhhmap}. However, under holographic duality, a negative $k$ is responsible to provide a real $U(1)$ charge of excited states and leave the vacuum charge to be pure imaginary. The negative $k$ leads to descendent states with negative norms, whose contribution to the partition function can be estimated and is much smaller than the primaries~\cite{Apolo:2018eky}. Furthermore, modular covariance requires that states with pure imaginary charges have to exist, consistent with the fact that global AdS$_3$ has pure imaginary charge.

In our previous paper, we find that the mean structure of WCFT, which contains a heavy-heavy-light three point function, can be estimated by a bulk tadpole diagram on a BTZ black hole background~\cite{Song:2019txa}. Since the operators have spin angular momentum, it turns out that the geodesic length approximation is not work directly. The propagator in the tadpole diagram is well approximated by a massive spinning particle's on-shell worldline action. For a general operator in WCFT with mass and spin, we propose that the bulk quantity that dual to the two point function for such operator is $e^{-S}$, where $S$ is the on-shell worldline action of a massive spinning particle with mass and spin related to the operator. The trajectory of the particle ends on the two points on the boundary. This works in a classical approximation with the range that the mass of the particle is much greater than 1 but still far less than $c$.

For the twist operators in WCFT, the scaling dimension and the $U(1)$ charge is given by \eqref{cdc}. For $n=1$, the twist operator should be trivially identity due to the normalization of the reduced density matrix. This is true by noticing that the vacuum charge are given by \eqref{vaccharge} and $c=3/(2G)$ in a holographic sense, and further requiring that $k=-1/G$ and $\alpha=\pi$. Compering to CFT on the algebraic level, $\tilde{L}_0$ and $-\tilde{P}_0$ play the role as left and right moving energies, respectively.
Near $n=1$, the twist operator is in classical approximation, and its dual object is a classical massive spinning particle with mass still in the perturbative region. In fact, by considering \eqref{dhhmap}, we have the following relation,
\begin{equation}\label{ms}
\Delta_n-\frac{Q_n^2}{k}+\frac{c}{24}=\frac{m_n+s_n}{2},~~~~-\frac{Q_n^2}{k}+\frac{c}{24}=\frac{m_n-s_n}{2}\,,
\end{equation}
where $m_n$ and $s_n$ are the mass and spin of the twist operator dual particle in the bulk AdS$_3$. The two point function of the twist operator near $n=1$ is now well approximated by
\begin{equation}\label{2pfcl}
\langle\Phi_n(u_1, v_1)\Phi_n^{\dagger}(u_2, v_2)\rangle_{\mathcal{C}}\sim e^{-S_{\mathrm{on-shell}}^{\mathrm{t}}}\,,
\end{equation}
where
\begin{equation}\label{toa}
S_{\mathrm{on-shell}}^{\mathrm{t}}=m_n\int_{(u_1, v_2)}^{(u_2, v_2)}d\tau+s_n\int_{(u_1, v_1)}^{(u_2, v_2)}d\tau\tilde{n}\cdot\nabla n\,,
\end{equation}
is the on-shell worldline action of a massive spinning particle with mass $m_n$ and spin $s_n$. From \eqref{ms}, using \eqref{cdc}, \eqref{vaccharge}, and $c=3/(2G)$, and further requiring $k=-1/G$ and $\alpha=\pi$, we find,
\begin{equation}\label{dn}
\partial_nm_n|_{n=1}=\frac{1}{8G},~~~~\partial_ns_n|_{n=1}=\frac{1}{8G}\,.
\end{equation}
According to \eqref{ee}, and using \eqref{2pfcl}, \eqref{toa}, \eqref{dn}, the entanglement entropy for a single interval with end points at $(u_1, v_1)$ and $(u_2,v_2)$ on the boundary can be expressed as
\begin{equation}
S_{\mathrm{EE}}=\frac{1}{8G}\left(\int_{(u_1, v_2)}^{(u_2, v_2)}d\tau+\int_{(u_1, v_1)}^{(u_2, v_2)}d\tau\tilde{n}\cdot\nabla n\right)\,.
\end{equation}
This can be viewed as the on-shell worldline action with mass and spin all equal to $1/(8G)$.  When $T_u=0$, setting $m=s=1/(8G)$ in \eqref{oa}, the entanglement entropy of a single interval with separation $(\Delta u, \Delta v)$ takes the form,
\begin{equation}\label{eebulk}
S_{\mathrm{EE}}=\frac{1}{4G}\left(T_v\Delta v+\log\frac{\Delta u}{\epsilon}\right)\,.
\end{equation}
This result also cover the case with $T_{u,v}=0$ by setting $m=s=1/(8G)$ in \eqref{oa0}. Under the holographic dictionary of the temperature obtained in~\cite{Song:2017czq},
\begin{equation}
T_u=\frac{\pi}{\beta},~~~~T_v=\frac{\bar{\beta}-\alpha}{\beta},
\end{equation}
and using \eqref{uvxy} and \eqref{L0P0}, and $k=-1/G$, the entanglement entropy from the bulk \eqref{eebulk} matches the entanglement entropy obtained by a pure field theory calculation $S_{\mathrm{EE}}=S_1$, with $S_1$ given in \eqref{eefield} in a zero temperature limit, i.e. $\beta\to\infty$, $\bar{\beta}\to\infty$, with $\bar{\beta}/\beta$ finite.

The bulk picture of a single interval entanglement entropy in WCFT here we present is different to the one studied in~\cite{Song:2016gtd}. In both pictures, the entangling surfaces in the bulk are all geodesics. The main differences come from two aspects: Geometrically, the geodesic line in~\cite{Song:2016gtd} is a straight line in the bulk without ending on the asymptotic boundary, while here the geodesic line is connected to the end points of the interval. And quantitatively, the quantity on the geodesic given in in~\cite{Song:2016gtd} that reflects the entanglement entropy is its length divided by $4G$, while here is the on-shell worldline action of a spinning particle. Both results are correct, and here we give an alternative way to find the gravity dual of the entanglement entropy in WCFT. Using the spinning particle's picture, it is interesting to study like the Witten diagrams or the R\'{e}nyi mutual information~\cite{Chen:2019xpb} in the context of AdS$_3$/WCFT, and we leave these as our future works.

\section*{Acknowledgement}
We are grateful to Luis Apolo, Alejandra Castro, Bin Chen, and Wei Song for helpful discussions. This work was partially supported by the Fundamental Research Funds for the Central Universities No. 2242019R10018.


\end{document}